\documentclass[11pt]{article}
\usepackage{pgf}
\renewcommand{\bibitem}{\vskip 2pt\par\hangindent\parindent\hskip-\parindent}

\title{``How many zombies do you know?''  Using indirect survey methods to measure  alien attacks and outbreaks of the undead}
\author{Andrew Gelman\footnote{Department of Statistics, Columbia University, New York.  Please do not tell my employer that I spent any time doing this.}\and
George A. Romero\footnote{Not really.}}

\date{12 Mar 2010}

\begin{document}
\maketitle
\begin{abstract}
The zombie menace has so far been studied only qualitatively or through the use of mathematical models without empirical content.  We propose to use a new tool in survey research to allow zombies to be studied indirectly without risk to the interviewers.
\end{abstract}

\baselineskip=18pt

\section{Introduction}

Zombification is a serious public-health and public-safety concern (Romero, 1968, 1978) but is difficult to study using traditional survey methods.  Zombies are believed to have very low rates of telephone usage and in any case may be reluctant to identify themselves as such to a researcher.  Face-to-face surveying involves too much risk to the interviewers, and internet surveys, although they originally were believed to have much promise, have recently had to be abandoned in this area because of the potential for zombie infection via computer virus.

\begin{figure}
\centerline{\includegraphics[width=3in]{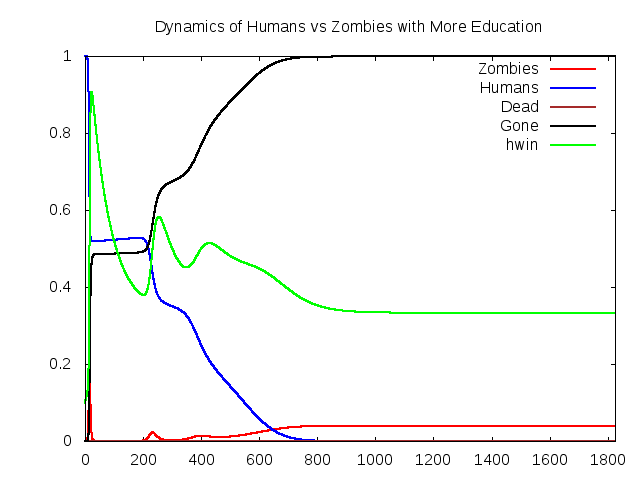}
\hspace{.5in}
\includegraphics[width=2in]{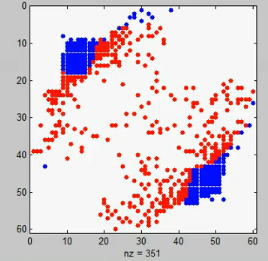}}
\caption{From Lakeland (2010) and Messer (2010).  There were other zombie graphs at these sites, but these were the coolest.}
\end{figure}

In the absence of hard data, zombie researchers\footnote{By ``zombie researchers,'' we are talking about people who research zombies.  We are not for a moment suggesting that these researchers are {\em themselves} zombies.  Just to be on the safe side, however, we have conducted all our interactions with these scientists via mail.} have studied outbreaks and their dynamics using differential equation models (Munz et al., 2009, Lakeland, 2010) and, more recently, agent-based models (Messer, 2010).  Figure 1 shows an example of such work.

But mathematical models are not enough.  We need data.

\section{Measuring zombification using network survey data}

Zheng, Salganik, and Gelman (2006) discuss how to learn about groups that are not directly sampled in a survey.  The basic idea is to ask respondents questions such as, ``How many people do you know named Stephen/Margaret/etc.'' to learn the sizes of their social networks, questions such as ``How many lawyers/teachers/police officers/etc. do you know,'' to learn about the properties of these networks, and questions such as ``How many prisoners do you know'' to learn about groups that are hard to reach in a sample survey.  Zheng et al.\ report that, on average, each respondent knows 750 people; thus, a survey of 1500 Americans can give us indirect information on about a million people.

This methodology should be directly applicable to zombies or, for that matter, ghosts, aliens, angels, and other hard-to-reach entities.  In addition to giving us estimates of the populations of these groups, we can also learn, through national surveys, where they are more prevalent (as measured by the residences of the people who know them), and who is more likely to know them.

A natural concern in this research is potential underreporting; for example, what if your wife\footnote{Here we are choosing a completely arbitrary example with absolutely no implications about our marriages or those of anyone we know.} is actually a zombie or an alien and you are not aware of the fact.  This bias can be corrected via extrapolation using the estimates of different populations with varying levels of reporting error; Zheng et al.\ (2006) discuss in the context of questions ranging from names (essentially no reporting error) to medical conditions such as diabetes and HIV that are often hidden.

\section{Discussion}

As Lakeland (2010) puts it, ``Clearly, Hollywood plays a vital role in educating the public about the proper response to zombie infestation.''  In this article we have discussed how modern survey methods based on social networks can help us estimate the size of the problem.

\begin{figure}
\centerline{\includegraphics[width=4.5in]{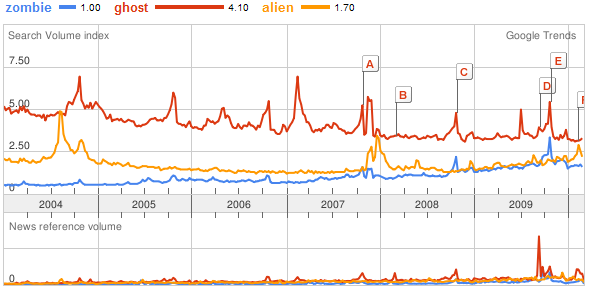}}
\caption{Google Trends report on ``zombie,'' ``ghost,'' and ``alien.''  The patterns show fascinating trends from which, we feel, much could be learned if resources were made available to us in the form of a sizable research grant from the Department of Defense, Department of Homeland Security, or a major film studio.  Please make out any checks to the first author or deposit directly to his PayPal account.}
\end{figure}

Other, related, approaches are worth studying too.  Social researchers have recently used Google Trends to study hard-to-measure trends using search volume (Askitas and Zimmerman, 2009, Goel, Hofman, et al., 2010); Figure 2 illustrates how this might be done in the zombie context.  It would also make sense to take advantage of social networking tools such as Facebook (Goel, Mason, et al., 2010) and more zombie-specific sites such as ZDate.  We envision vast unfolding vistas of funding in this area.

\section{Technical note}

We originally wrote this article in Word, but then we converted it to Latex to make it look more like science.

\section{References}

\noindent

\bibitem Askitas, N., and Zimmermann, K. F. (2009).  Google
econometrics and unemployment forecasting.  {\em Applied
Economics Quarterly} {\bf 55}, 107--120.

\bibitem Goel, S., Hofman, J. M., Lahaie, S., Pennock, D. M., and Watts, D. J. (2010).  What can search predict.  Technical report, Yahoo Research.

\bibitem Goel, S., Mason, W., and Watts, D. J. (2010).  Real and perceived attitude homophily in social networks.  Technical report, Yahoo Research.

\bibitem Lakeland, D. (2010).  Improved zombie dynamics.  Models of Reality blog, 1 March.  {\tt http://models.street-artists.org/?p=554}

\bibitem Messer, B. (2010).  Agent-based computational model of humanity's prospects for post zombie outbreak survival.  The Tortise's Lens blog, 10 March.\newline
 {\tt http://thetortoiseslens.blogspot.com/2010/03/}\newline
{\tt agent-based-computational-model-of.html}

\bibitem Munz, P., Hudea, I., Imad, J., and Smith?, R. J. (2009).  When zombies attack!:  Mathematical modelling of an outbreak of zombie infection.  In {\em Infectious Disease Modelling Research Progress}, ed. J. M. Tchuenche and C. Chiyaka, 133--150.  Hauppage, New York:  Nova Science Publishers.

\bibitem Romero, G. A. (1968).  {\em Night of the Living Dead}.  Image Ten.

\bibitem Romero, G. A. (1978).  {\em Dawn of the Dead}.  Laurel Group.

\bibitem Zheng, T., Slaganik, M., and Gelman, A. (2006).  ``How many people do you know in prison?'':  Using overdispersion in count data to estimate social structure in networks. {\em Journal of the American Statistical Association} {\bf 101}, 409--423.

\end{document}